\documentclass[fleqn,11pt]{article}

\usepackage{latexsym,ifthen,epsfig,color}
\usepackage{amsmath,amssymb,amsthm}
\usepackage{float}

\newcommand{\NP}{\ensuremath{\mathbb{NP}}}

\newtheorem{theorem}{Theorem}
\newtheorem{lemma}{Lemma}
\newtheorem{corollary}{Corollary}

\newtheorem{claim}{Claim}

\begin{document}

\title{Weighted Efficient Domination for $(P_5+kP_2)$-Free Graphs in Polynomial Time}

\author{
Andreas Brandst\"adt\footnote{Fachbereich Informatik, 
Universit\"at Rostock, A.-Einstein-Str. 22, D-18051 Rostock, Germany.
\texttt{e-mail: ab@informatik.uni-rostock.de}
}
\and
Vassilis Giakoumakis\footnote{MIS (Mod\'elisation, Information \& Syst\`emes), Universit\'e de Picardie
Jules Verne, Amiens, France.
\texttt{e-mail: vassilis.giakoumakis@u-picardie.fr}
}
}

\maketitle

\begin{abstract}
Let $G$ be a finite undirected graph.
A vertex {\em dominates} itself and all its neighbors in $G$.
A vertex set $D$ is an {\em efficient dominating set} (\emph{e.d.}\ for short) of $G$ if every vertex of $G$ is dominated by exactly one vertex of $D$.
The \emph{Efficient Domination} (ED) problem, which asks for the existence of an e.d.\ in $G$, is known to be \NP-complete even for very restricted graph classes such as for claw-free graphs, for chordal graphs and for $2P_3$-free graphs (and thus, for $P_7$-free graphs). We call a graph $F$ a {\em linear forest} if $F$ is cycle- and claw-free, i.e., its components are paths. Thus, the ED problem remains \NP-complete for $F$-free graphs, whenever $F$ is not a linear forest. Let WED denote the vertex-weighted version of the ED problem asking for an e.d. of minimum weight if one exists.

In this paper, we show that WED is solvable in polynomial time for $(P_5+kP_2)$-free graphs for every fixed $k$, which solves an open problem, and, using modular decomposition, we improve known time bounds for WED on $(P_4+P_2)$-free graphs, $(P_6,S_{1,2,2})$-free graphs, and on $(2P_3,S_{1,2,2})$-free graphs and simplify proofs. For $F$-free graphs, the only remaining open case is WED on $P_6$-free graphs. 
\end{abstract}


\section{Introduction}\label{sec:intro}

Let $G=(V,E)$ be a finite undirected graph. A vertex $v \in V$ {\em dominates} itself and its neighbors. A vertex subset $D \subseteq V$ is an {\em efficient dominating set} ({\em e.d.} for short) of $G$ if every vertex of $G$ is dominated by exactly one vertex in $D$.
Note that not every graph has an e.d.; the {\sc Efficient Dominating Set} (ED) problem asks for the existence of an e.d.\ in a given graph $G$. 
If a vertex weight function $\omega: V \to \mathbb{N}$ is given, the {\sc Weighted Efficient Dominating Set} (WED) problem asks for a minimum weight e.d. in $G$ if there is one $G$ or for determining that $G$ has no e.d.

For a set ${\cal F}$ of graphs, a graph $G$ is called {\em ${\cal F}$-free} if $G$ contains no induced subgraph isomorphic to a member of ${\cal F}$.
For two graphs $F$ and $G$, we say that $G$ is $F$-free if $G$ is $\{F\}$-free.
We denote by $G+H$ the disjoint union of graphs $G$ and $H$.
Let $P_k$ denote a chordless path with $k$ vertices, and let $2P_k$ denote $P_k+P_k$, and correspondingly for $kP_2$. The {\it claw} is the $4$-vertex tree with three vertices of degree $1$. 

Many papers have studied the complexity of ED on special graph classes - see e.g. \cite{BraMilNev2013} for references. In particular, ED remains \NP-complete for $2P_3$-free graphs, for chordal graphs, for line graphs and thus for claw-free graphs. 

A {\it linear forest} is a graph whose components are paths; equivalently, it is a graph which is cycle-free and claw-free. 
The \NP-completeness of ED on chordal graphs and on claw-free graphs implies: If $F$ is not a linear forest, then ED is \NP-complete on $F$-free graphs.
This motivates the analysis of ED/WED on $F$-free graphs for linear forests $F$.

\medskip

In this paper, we show that WED is solvable in polynomial time for $(P_5+kP_2)$-free graphs for every fixed $k$, which solves an open problem, and, using modular decomposition, we improve known time bounds for WED on $(P_4+P_2)$-free graphs, $(P_6,S_{1,2,2})$-free graphs, and on $(2P_3,S_{1,2,2})$-free graphs and simplify proofs (see \cite{BraMilNev2013,Nevri2014} for known results). For $F$-free graphs, the only remaining open case is WED on $P_6$-free graphs. 

Various of our algorithms are robust in the sense of \cite{Spinr2003}, that is, a robust algorithm for a graph class ${\cal C}$ works on every input graph $G$ and either solves the problem correctly or states that $G \not \in {\cal C}$. We say that the algorithm is {\em weakly robust} if it either gives the optimal WED solution for the input graph $G$ or states that $G$ has no e.d. or is not in the class.  

\section{Basic Notions and Results}\label{sec:basicnotions}

\subsection{Some Basic Notions}

All graphs considered in this paper are finite, undirected and simple (i.e., without loops and multiple edges). For a graph $G$, let $V(G)$ or simply $V$ denote its vertex set and $E(G)$ or simply $E$ its edge set; throughout this paper, let $|V|=n$ and $|E|=m$. We can assume that $G$ is connected (otherwise, WED can be solved separately for its components); thus, $m \ge n-1$. 
A graph is {\em nontrivial} if it has at least two vertices.
For a vertex $v \in V$, $N(v)=\{u \in V \mid uv \in E\}$ denotes its ({\em open}) {\em neighborhood}, and $N[v]=\{v\} \cup N(v)$ denotes its {\em closed neighborhood}. A vertex $v$ {\em sees} the vertices in $N(v)$ and {\em misses} all the others. The {\em anti-neighborhood} of vertex $v$ is $A(v)=V \setminus N[v]$.

For a vertex set $U \subseteq V$, its neighborhood is $N(U)=\{x \mid x \notin U, \exists y \in U, xy \in E\}$, and its anti-neighborhood $A(U)$ is the set of all vertices not in $U$ missing $U$.

The {\em degree} of a vertex $x$ in a graph $G$ is $d(x):= |N(x)|$. Let $\delta(G)$ denote the minimum degree of any vertex in $G$.

A vertex $u$ is {\em universal} for $G=(V,E)$ if $N[u]=V$.
Independent sets, complement graph, and connected components are defined as usual.
%The \emph{girth} of a graph with at least one cycle is the shortest length of its cycle.
Unless stated otherwise, $n$ and $m$ will denote the number of vertices and edges, respectively, of the input graph.

\subsection{A General Approach for the WED Problem}

For a graph $G=(V,E)$ and a vertex $v \in V$, the {\em distance levels} with respect to $v$ are $$N_i(v) = \{ w \in V \mid dist(v,w)=i \}$$ for all $i \in N$.
If $v$ is fixed, we denote $N_i(v)$ by $N_i$. Let $R:=V \setminus (\{v\} \cup N_1 \cup N_2)$, and let $G_v:=G[N_2 \cup R]$ where vertices in $N_2$ get weight $\infty$. 
Obviously, we have: $G$ has a finite weight e.d. $D_v$ with $v \in D_v$ if and only if $G_v$ has a finite weight e.d. $D$, and $D_v=\{v\} \cup D$. 
In some cases, for every vertex $v \in V$, the WED problem can be efficiently solved on $G_v$, say in time $t(m)$ with $t(m) \ge m$. 

If graph $G=(V,E)$ has an e.d. $D$ then for any vertex $v \in V$, either $v \in D$ or one of its neighbors is in $D$. 
Thus, if $deg_G(v)=\delta(G)$, one has to consider the WED problem on $G_x$ for $\delta(G)+1$ vertices $x \in N[v]$. Thus we obtain:

\begin{lemma}\label{WEDG_v}
If for a graph class ${\cal C}$ and input graph $G=(V,E)$ in ${\cal C}$, WED is solvable in time $t(m)$ on $G_v$ for all $v \in V$ then WED is solvable in time $O(\delta(G) \cdot t(m))$ for graph class ${\cal C}$.
\end{lemma} 

\subsection{Linear Forests}

%The approach above implies: 
%\begin{lemma}\label{lma:nearly}
%If WED is solvable in polynomial time on $F$-free graphs, then WED is solvable in polynomial time on $(F+P_1)$-free graphs.
%\end{lemma}

As already mentioned, if $F$ is a linear forest such that one of its components contain $2P_3$, or two of its components contain $P_3$, the WED problem is \NP-complete for $F$-free graphs. 

For $2P_2$-free graphs and more generally, for $kP_2$-free graphs, it is known that the number of maximal independent sets is polynomial \cite{BalYu1989,Farbe1989,Prisn1995} and can be enumerated efficiently \cite{TsuIdeAriShi1977}. Since every e.d. is a maximal independent set, WED can be solved in polynomial time for $kP_2$-free graphs.

In Section \ref{P5P2free}, we show that WED is solvable in polynomial time for $(P_5+kP_2)$-free graphs for every fixed $k$. 
Thus, the only remaining open case is the one of $P_6$-free graphs; our approach used for $(P_5+kP_2)$-free graphs shows that if WED is polynomial for $P_6$-free graphs then it is polynomial for $(P_6+kP_2)$-free graphs as well.    

\subsection{Modular Decomposition for the WED Problem}
\label{sec:modular}

A set $H$ of at least two vertices of a graph $G$ is called \emph{homogeneous} if $H \not= V(G)$ and every vertex outside $H$ is either adjacent to all vertices in $H$, or to no vertex in $H$. Obviously, $H$ is homogeneous in $G$ if and only if $H$ is homogeneous in the complement graph $\overline{G}$.
A graph is {\em prime} if it contains no homogeneous set. A homogeneous set $H$ is \emph{maximal} if no other homogeneous set properly contains $H$.
It is well known that in a connected graph $G$ with connected complement $\overline{G}$, the maximal homogeneous sets are pairwise disjoint and can be determined in linear time using the so called {\it modular decomposition} (see, e.g., \cite{McCSpi1999}). The {\em characteristic graph} $G^*$ of $G$ is the graph obtained from $G$ by contracting each of the maximal homogeneous sets $H$ of $G$ to a single representative vertex $h \in H$, and connecting two such vertices by an edge if and only if they are adjacent in $G$. It is well known (and can be easily seen) that $G^*$ is a prime graph.

For a disconnected graph $G$, the WED problem can be solved separately for each component. If $\overline{G}$ is disconnected, then obviously, $D$ is an e.d. of $G$ if and only if $D$ is a single universal vertex of $G$. Thus, from now on, we can assume that $G$ and $\overline{G}$ are connected, and thus, maximal homogeneous sets are pairwise disjoint. Obviously, we have:

\begin{lemma}\label{edhom}
Let $H$ be a homogeneous set in $G$ and $D$ be an e.d. of $G$. Then the following properties hold:
\begin{itemize}
\item[$(i)$] $|D \cap H| \le 1$. 
\item[$(ii)$] If $H$ has no vertex which is universal for $H$ then $|D \cap H|=0$.  
\end{itemize}
\end{lemma} 

Thus, the WED problem on a connected graph $G$ for which $\overline{G}$ is connected can be easily reduced to its characteristic graph $G^*$ by contracting each homogeneous set $H$ to a single representative vertex $h$ whose weight is either $\infty$ if $H$ has no universal vertex or the minimum weight of a universal vertex in $H$ otherwise. Obviously, $G$ has an e.d. $D$ of finite weight if and only if $G^*$ has a corresponding e.d. of the same weight. Thus, we obtain:  

\begin{theorem}\label{thm:prime}
Let ${\cal G}$ be a class of graphs and ${\cal G}^*$ the class of all prime induced subgraphs of the graphs in ${\cal G}$. If the (W)ED problem can be solved for
graphs in ${\cal G}^*$ with $n$ vertices and $m$ edges in time $O(T(n,m))$, then the same problem can be solved for graphs in ${\cal G}$ in time $O(T(n,m)+m)$.
\end{theorem}

The modular decomposition approach leads to a linear time algorithm for WED on $2P_2$-free graphs (see \cite{BraMilNev2013}) and to a very simple $O(\delta(G) m)$ time algorithm for WED on $P_5$-free graphs (a simplified variant of the corresponding result in \cite{BraMilNev2013}); the modular decomposition approach is also described in \cite{Nevri2014}. 

\section{WED in Polynomial Time for $(P_5+kP_2)$-Free Graphs}\label{P5P2free}

In this section we solve an open problem from \cite{BraMilNev2013}. 
Let $G$ be a $(P_5+P_2)$-free graph and assume that $G$ is not $P_5$-free; otherwise, WED can be solved in time $O(\delta(G) m)$ as described in \cite{BraMilNev2013}. 
Let $v_1,v_2,v_3,v_4,v_5$ induce a $P_5$ $H$ in $G$ with edges $v_1v_2,v_2v_3,v_3v_4,v_4v_5$, let $X=N(H)=\{x \mid x \notin V(H), \exists i(xv_i \in E)\}$ denote the neighborhood of $H$ and let $Y$ denote the anti-neighborhood $A(H)$ of $H$ in $G$. Since $G$ is $(P_5+P_2)$-free, we have: 

\begin{claim}\label{Ystable} 
$Y$ is an independent set. 
\end{claim}

Assume that $G$ has an e.d. $D$.  Then:

\begin{claim}\label{XD5} 
$|(V(H) \cup X) \cap D| \le 5$.
\end{claim}

\noindent
{\em Proof of Claim $\ref{XD5}$.}
Obviously, $|V(H) \cap D| \le 2$. If $|V(H) \cap D| = 2$ then $|X \cap D| = 0$ or $|X \cap D| = 1$ since $D$ is an e.d. If $|V(H) \cap D| = 1$ then $|X \cap D| \le 3$. Finally, if
$|V(H) \cap D| = 0$ then $|X \cap D| \le 5$ which shows Claim $\ref{XD5}$. 
$\Diamond$

\medskip

Let $D = D_1 \cup D_2$ be the partition of $D$ into $D_1= D \cap (V(H) \cup X)$ and $D_2= D \cap Y$. 

\begin{claim}\label{Dpartition} 
$D_2 = A(D_1) \cap Y$.
\end{claim}

\noindent
{\em Proof of Claim $\ref{Dpartition}$.}
Since the anti-neighborhood $Y$ of $H$ is an independent set, every vertex in $Y$ can only be dominated by itself or by a vertex from $D \cap X$. Thus, Claim    
$\ref{Dpartition}$ holds.
$\Diamond$

\medskip

This leads to the following simple algorithm for checking whether $G$ has an e.d. $D$: 

\begin{enumerate}
\item Check whether $G$ is $P_5$-free; if yes, apply the corresponding algorithm for WED on $P_5$-free graphs (which works in time $O(\delta(G) m)$), otherwise let $H$ be a $P_5$ in $G$. Determine $X=N(H)$ and $Y=A(H)$. If $Y$ is not independent then $G$ is not $(P_5+P_2)$-free. Otherwise do the following: 
\item For every independent set $S \subseteq V(H) \cup X$ with $|S| \le 5$, check whether $S \cup (A(S) \cap Y)$ is an e.d.
\item If there is such a set then take one of minimum weight, otherwise output ``$G$ has no e.d.''.  
\end{enumerate}

Obviously, the algorithm is correct and its running time is at most $O(n^5 m)$.

\medskip

For every fixed $k$, the approach for $(P_5+P_2)$-free graphs can be generalized to $(P_5+kP_2)$-free graphs:  
Assume inductively that WED can be solved in polynomial time for $(P_5+(k-1)P_2)$-free graphs. Thus, if the given graph $G$ is $(P_5+(k-1)P_2)$-free, we can use the assumption, otherwise find (in polynomial time) an induced subgraph $H$ isomorphic to $P_5+(k-1)P_2$ and determine its neighborhood $X$ and its anti-neighborhood $Y$. Then similar claims as in the $(P_5+P_2)$-free case hold; in particular, $Y$ is independent, $|(V(H) \cup X) \cap D| \le 5+2k$ and we can check whether for such an independent set $S$ and partition $D=D_1 \cup D_2$, $D_2 = A(D_1) \cap Y$ holds. 

\begin{corollary}\label{EDP5+kP2freegr} 
For every fixed $k$, WED is solvable in polynomial time for $(P_5+kP_2)$-free graphs.
\end{corollary}

The approach can be easily generalized to $(H+kP_2)$-free graphs whenever WED is solvable in polynomial time for $H$-free graphs. However, WED remains \NP-complete for $(H+kP_2)$-free graphs whenever WED is \NP-complete for $H$-free graphs. If WED is solvable in polynomial time for $P_6$-free graphs then it is  solvable in polynomial time for$(P_6+kP_2)$-free graphs for every fixed~$k$.   

\section{WED for $(P_4+P_2)$-Free Graphs in Time $O(\delta(G) m)$}\label{P4+P2free}

In this section we slightly improve the time bound $O(nm)$ for WED \cite{BraMilNev2013} to $O(\delta(G) m)$ and simplify the proof in  \cite{BraMilNev2013}.  
According to Lemma \ref{WEDG_v}, for a vertex $v \in V$ with minimal degree $\delta(G)$, we check for all $x \in N[v]$ whether $G_x$ has an e.d. $D_x$.  We first collect some properties assuming that $G$ is $(P_4+P_2)$-free and has an e.d. $D_v$. As before, let $G_v:=G[N_2 \cup R]$; we can assume that $G_v$ is prime. 
We are looking for an e.d. of $G_v$ with finite weight and assume that $D_v \setminus \{v\}$ is such an e.d.  
Since $G$ is $(P_4+P_2)$-free, we have: 

\begin{claim}\label{Rcograph} 
$G[R]$ is a cograph. 
\end{claim}

Let $R_1,\ldots,R_{\ell}$ denote the connected components of $G[R]$. Note that an e.d. of a connected cograph $H$ has only one vertex, namely a universal vertex of $H$. Thus:

\begin{claim}\label{cographeduniv} 
For all $i \in \{1,\ldots,\ell\}$, $|D_v \cap R_i|=1$, and in particular, if $d \in D_v \cap R_i$ then $d$ is universal for $R_i$. 
\end{claim}

For all $i \in \{1,\ldots,\ell\}$, let $D_v \cap R_i =\{d_i\}$. 
Let $U_i$ be the set of universal vertices in $R_i$. Thus, if $U_i=\emptyset$ then $G$ has no e.d., and if $U_i=\{d_i\}$ then necessarily $d_i \in D_v$.
From now on, assume that for every $i \in \{1,\ldots,\ell\}$, $|U_i| \ge 2$. We first claim that $\ell>1$: Since $G_v$ is prime and in case $\ell=1$, $G_v$ has an e.d. (of finite weight) if and only if $G_v$ contains a universal vertex $z \in R_1$ for $G_v$, it follows:

\begin{claim}\label{ell>1} 
For prime $G_v$ with e.d. $D_v$ of finite weight, $\ell>1$ holds. 
\end{claim}

Since for finding an e.d. in $G_v$, every $R_i$ can be reduced to the set $U_i$ of its universal vertices (since the non-universal vertices in $R_i$ cannot dominate all $R_i$ vertices), we can assume that for all $i \in \{1,\ldots,\ell\}$, $R_i$ is a clique. If $|N_2| = 1$ then, since $G_v$ is prime, for all $i \in \{1,\ldots,\ell\}$, $|R_i| \le 2$ and thus, $G_v$ is a tree (in particular: If there are $i,j \in \{1,\ldots,\ell\}$, $i \neq j$, $|R_i| = |R_j| =1$ then $G_v$ has no e.d., if there is exactly one $i \in \{1,\ldots,\ell\}$ with $|Z_i|=1$ then this determines the $D_v$ vertices in $Z$, and if for all $i \in \{1,\ldots,\ell\}$, $|R_i| = 2$ then one has to choose the e.d. with smallest weight in the obvious way). From now on, let $|N_2| \ge 2$. If for all $z \in R$, either $z$ has a join or a co-join to $N_2$ then $N_2$ would be homogeneous in $G_v$ - contradiction. Thus, from now on we have: 

\begin{claim}\label{neighbnonneighb} 
There is a vertex $z \in R$ having a neighbor and a non-neighbor in $N_2$. 
\end{claim}

Since $G$ is $(P_4+P_2)$-free, we have: 

\begin{claim}\label{atmost1nonn} 
If $x \in N_2$ has a neighbor in $R_i$ then for all $j \neq i$, it has at most one non-neighbor in $R_j$.   
\end{claim}

In particular, this means: 

\begin{claim}\label{forced1forcedall} 
If $x \in N_2$ is adjacent to $d_i \in R_i \cap D_v$ then for all $j \neq i$, $x$ has exactly one non-neighbor in $Z_j$ which is the $D_v$-vertex in $R_j$.
\end{claim}

\begin{claim}\label{znonneighb} 
If a vertex $z \in R_i$ has a non-neighbor $x \in N_2$ and $z \not\in D_v$ then for all $j \neq i$, $x$ has exactly one non-neighbor in $R_j$, namely $xd_j \notin E$ for $d_j \in R_j \cap D_v$.
\end{claim}

\noindent
{\em Proof of Claim $\ref{znonneighb}$.} 
Let $z \in R_1$ have non-neighbor $x \in N_2$, and $z \notin D_v$, i.e., $z \neq d_1$. Then, since $G$ is $(P_4+P_2)$-free, $xd_1 \in E$. By Claim \ref{forced1forcedall}, $x$ has exactly one non-neighbor in $R_j$ for each $j \in \{2,\ldots,\ell\}$ (which is the corresponding $D_v$ vertex in $R_j$).  
\qed

\medskip

{\bf Algorithm $(P_4+P_2)$-Free-WED-$G_v$:} 

\medskip

\noindent
{\bf Given:} Graph $G=(V,E)$ and prime graph $G_v=G[N_2 \cup R]$ as constructed above with vertex weights $w(x)$; for all $x \in N_2$, $w(x)=\infty$.\\ 
{\bf Output:} An e.d. $D_v$ of $G_v$ of finite minimum weight, if $G_v$ has an e.d., or the statement that $G$ is not $(P_4+P_2)$-free or $G_v$ does not have any e.d. of finite weight.

\begin{enumerate}
\item[(0)] Initially, $D_v:=\emptyset$. 
\item[(1)] Check if $G[R]$ is a cograph. If not then $G$ is not $(P_4+P_2)$-free - STOP. Else determine the connected components $R_1,\ldots,R_{\ell}$ of $G[R]$. If $\ell=1$ then $G_v$ has no e.d. of finite weight - STOP.
\item[(2)] For all $i \in \{1,\ldots,\ell\}$, determine the set $U_i$ of universal vertices in $R_i$. If for some $i$, $U_i=\emptyset$ then $G_v$ has no e.d. - STOP. From now on, let $R_i:=U_i$. If $U_i=\{d_i\}$ then $D_v:=D_v \cup \{d_i\}$.
\item[(3)] If $|N_2|=1$ then check whether $G_v$ is a tree and solve the problem in the obvious way. If $|N_2| > 1$, choose a vertex $z \in R$ with a neighbor $w \in N_2$ and a non-neighbor $x \in N_2$, say $z \in R_i$. 
\begin{enumerate}
\item[(3.1)] Check if $z \in D_v$ leads to an e.d. (by using neighbor $w$ of $z$ and Claim \ref{forced1forcedall}).
\item[(3.2)] Check if $z \notin D_v$ leads to an e.d. (by using the non-neighbors of $x$ in $R_j$, $j \neq i$ and Claim \ref{znonneighb})  
\item[(3.3)] If there is no e.d. in both cases (3.1) and (3.2) then either $G$ is not $(P_4+P_2)$-free or has no e.d. of finite weight - STOP.
\end{enumerate}
\end{enumerate}

\begin{theorem}\label{correcttime} 
Algorithm $(P_4+P_2)$-Free-WED-$G_v$ is correct and runs in time $O(m)$.
\end{theorem}

\noindent  
{\bf Proof.} 
{\em Correctness.} The correctness follows from Claims \ref{Rcograph} - \ref{znonneighb}.

\medskip

\noindent
{\em Time bound.} The linear time bound is obvious.
\qed

\begin{corollary}\label{WEDP4+P2} 
WED is solvable in time $O(\delta(G) m)$ for $(P_4+P_2)$-free graphs.
\end{corollary}
  
\section{WED for Some Subclasses of $P_6$-Free Graphs}\label{P6free}
 
Recall that the complexity of WED for $P_6$-free graphs is open. In this section we consider WED for some subclasses of $P_6$-free graphs.
Let $G=(V,E)$ be a prime $P_6$-free graph, let $v \in V$ and let $N_1,N_2,\dots$ be the distance levels of $v$. Then we have: 
\begin{equation}\label{N5emptyN4stable}
N_k = \emptyset \mbox{ for all } k \ge 5 \mbox{ and } N_4 \mbox{ is an independent vertex set}.
\end{equation}

 Assume that $G$ admits an e.d.\ $D_v$ of finite weight with $v \in D_v$. Let $G_v:=G[N_2 \cup N_3 \cup N_4]$; we can assume that $G_v$ is prime. 
As before, $D_v \cap (N_1 \cup N_2) = \emptyset$; set $w(x)=\infty$ for $x \in N_2$. Thus, vertices of $N_2$ have to be dominated by vertices of $D_v \cap N_3$. We claim:
\begin{equation}\label{atmostoneDN3N4P6}
\mbox{At most one vertex in } D_v \cap N_3 \mbox{ has neighbors in } N_4.
\end{equation}

{\em Proof.}
Assume that there are two vertices $d_1,d_2 \in N_3 \cap D_v$ with neighbors in $N_4$, say $x_i \in N_4$ with $d_ix_i \in E$ for $i=1,2$.
Let $b_i \in N_2$ with $b_id_i \in E$ for $i=1,2$.
Since $D_v$ is an e.d., $b_1 \neq b_2$ and $x_1 \neq x_2$ and $d_1$ misses $b_2,x_2$ while $d_2$ misses $b_1,x_1$.
Since $N_4$ is independent, $x_1x_2 \notin E$ holds. 
Now, if $b_1b_2 \in E$, then $x_1,d_1,b_1,b_2,d_2,x_2$ induce a $P_6$ in $G$, and if $b_1b_2 \notin E$, there is a $P_6$ as well (together with $N_1$ vertices), a contradiction.
\qed
 
\subsection{WED for $(P_6,S_{1,2,2})$-free graphs in time $O(\delta(G) m)$}\label{P6S122free}

In this subsection we improve the time bound $O(n^2m)$ for WED \cite{BraMilNev2013} to $O(\delta(G) m)$ and simplify the proof in \cite{BraMilNev2013}.  
Let $G=(V,E)$ be a connected $(P_6,S_{1,2,2})$-free graph, let $v \in V$ and let $N_1,N_2,\dots$ be the distance levels of $v$.
We claim:
\begin{equation}\label{DcapN4empty}
D_v \cap N_4 = \emptyset.
\end{equation}
{\em Proof.}
Assume to the contrary that there is a vertex $d \in D_v \cap N_4$.
Let $c \in N_3$ be a neighbor of $d$, let $b \in N_2$ be a neighbor of $c$ and let $a \in N_1$ be a neighbor of $b$.
Then $b$ has to be dominated by a $D_v$-vertex $d' \in N_3$, and since $D_v$ is an e.d., $cd' \notin E$ and $dd' \notin E$
but now, $v,a,b,c,d,d'$ induce an $S_{1,2,2}$, a contradiction.
\qed

\medskip

Thus, set $w(x):=\infty$ for all $x \in N_4$. By (\ref{DcapN4empty}), $D_v \subseteq N_3 \cup \{v\}$. 
Claim (\ref{atmostoneDN3N4P6}) means that $D_v$ vertices in $N_3$ have either a join or a co-join to $N_4$. Thus for finding an e.d. of $G_v$, we can delete all vertices in $N_3$ which have a neighbor and a non-neighbor in $N_4$. Reducing $G_v$ in this way gives $G_v'$; 
again, we can assume that $G_v'$ is prime. Now, $N_4$ is a module and thus, $|N_4| \le 1$. Let $N_4=\{z\}$ if $N_4$ is nonempty. Let $Q_1,\ldots, Q_{\ell}$ denote the connected components of $G[N_3]$. We claim:
\begin{equation}\label{onlyoneDinN3}
\mbox{No component } Q_i \mbox{ in } G[N_3] \mbox{ contains two vertices of } D_v.
\end{equation}
{\em Proof.}
Assume to the contrary that $Q_1$ contains $d_1,d_2 \in D_v$, $d_1 \neq d_2$.
Let $x \in N_2$ be a neighbor of $d_1$, and let $P$ denote a path in $Q_1$ connecting $d_1$ and $d_2$, i.e., either $P=(d_1,x_1,x_2,d_2)$ or $P=(d_1,x_1,x_2,x_3,d_2)$. Let $a$ be a common neighbor of $x$ and $v$. If $P=(d_1,x_1,x_2,d_2)$ then $x$ is not adjacent to $x_2$ since $G$ is $S_{1,2,2}$-free (otherwise $v,a,x,d_1,x_2,d_2$ induce an $S_{1,2,2}$) and since $G$ is $P_6$-free, $x$ is adjacent to $x_1$ (otherwise $v,a,x,d_1,x_1,x_2$ induce a $P_6$) but now $v,a,x,x_1,x_2,d_2$ induce a $P_6$ - contradiction. If $P=(d_1,x_1,x_2,x_3,d_2)$, the arguments are similar. 
\qed

\medskip

Thus, by (\ref{onlyoneDinN3}), if $D_v$ is an e.d. of $G_v$ then $|D_v \cap Q_i|=1$ for all $i$, $1 \le i \le \ell$, and the corresponding $D_v$-vertex is universal for $Q_i$. Thus, we can restrict $Q_i$ to its universal vertices $U_i$ (which means that now, $Q_i$ is a clique; if $U_i=\emptyset$ then $G_v$ has no e.d.) 
In case $\ell=1$ this means that if $G_v$ has an e.d., $G_v$ must have a universal vertex (since a $D_v$-vertex being universal for $Q_1$ must also be universal for $N_2 \cup N_4$) which is impossible since $G_v$ is prime. This implies $\ell>1$. If $|Q_i|=1$ then the corresponding vertex in $Q_i$ is a forced vertex for $D_v$ and has to be added to $D_v$. We claim:
\begin{equation}\label{onlyonedistingN2}
N_2 \mbox{ vertices cannot distinguish more than one } Q_i, i \in \{1,\ldots,\ell\}.
\end{equation}
{\em Proof.}
Since $G$ is $S_{1,2,2}$-free, no vertex in $N_2$ can distinguish two components $Q_i,Q_j$ in $N_3$. In order to show (\ref{onlyonedistingN2}), assume to the contrary that there are components $Q_1,Q_2$ in $G_v'$ with $c_1,d_1 \in Q_1$ and $c_2,d_2 \in Q_2$ which are distinguished by vertices $x_1,x_2 \in N_2$ such that $x_1d_1 \in E$, $x_1c_1 \notin E$, and $x_2d_2 \in E$, $x_2c_2 \notin E$. Since no vertex in $N_2$ can distinguish two components $Q_i,Q_j$, $x_1 \neq x_2$ holds, and since $G$ is $S_{1,2,2}$-free, $x_1c_2 \notin E$ and $x_1d_2 \notin E$, and by symmetry also $x_2c_1 \notin E$ and $x_2d_1 \notin E$, but now $c_1,d_1,x_1,x_2,d_2,c_2$ induce a $P_6$ if $x_1x_2 \in E$ or a $P_6$ together with $N_1$ vertices if $x_1x_2 \not\in E$ - contradiction. 
\qed

\medskip

First assume $N_4 \neq \emptyset$, i.e., $N_4=\{z\}$.
For every $i \in \{1,\ldots,\ell\}$, let $Q_i^+$ denote the neighbors of $z$ in $Q_i$ and let $Q_i^-$ denote the non-neighbors of $z$ in $Q_i$. 
By (\ref{onlyonedistingN2}), at most one $Q_i$ has more than two vertices, say $|Q_i| \le 2$ for all $i \in \{2,\ldots,\ell\}$ since in this case, $Q_i^+$ and $Q_i^-$ are modules. Since $D_v$ is an e.d., there is a vertex $d \in D_v$ with $dz \in E$; say $d \in Q_i^+$. Let $b \in N_2$ with $bd \in E$ and $a \in N_1$ with $ab \in E$. Now for $j \neq i$, every neighbor $x \in Q_j^+$ of $z$ must see $b$ since otherwise $v,a,b,d,z,x$ induce a $P_6$, and every non-neighbor $y \in Q_j^-$ of $z$ must miss $b$ since otherwise $v,a,b,d,z,y$ induce an $S_{1,2,2}$ but if $Q_j$ contains both $x$ and $y$ then $v,a,b,d,x,y$ induce an $S_{1,2,2}$ - contradiction.
Thus, we have:
\begin{equation}\label{atmostonemorethannone}
\mbox{ At most one } Q_i \mbox{ has more than one vertex}. 
\end{equation}
Say $|Q_i| = 1$ for all $i \in \{2,\ldots,\ell\}$. If $N_4=\emptyset$, this holds as well. 

\medskip

This leads to the following algorithm for WED with time bound $O(m)$ for every $v$:

\medskip

{\bf Algorithm $(P_6,S_{1,2,2})$-Free-WED-$G_v$:} 

\medskip

\noindent
{\bf Given:} Connected graph $G=(V,E)$ and prime graph $G_v=G[N_2 \cup N_3 \cup N_4]$ as constructed above with vertex weights $w(x)$; for all $x \in N_2 \cup N_4$, $w(x)=\infty$. \\ 
{\bf Output:} An e.d. $D_v$ of $G_v$ of finite weight, if $G_v$ has such an e.d., or the statement that $G$ is not $(P_6,S_{1,2,2})$-free or $G_v$ does not have any e.d. of finite weight.

\begin{enumerate}
\item[(0)] Initially, $D_v:=\emptyset$.
\item[(1)] Check if $G[N_5]=\emptyset$; if not then $G$ is not $P_6$-free - STOP. Else determine the connected components $Q_1,\ldots,Q_{\ell}$ of $G[N_3]$. If $\ell=1$ then $G_v$ has no e.d. of finite weight - STOP.
\item[(2)] For all $i \in \{1,\ldots,\ell\}$, determine the set $U_i$ of universal vertices in $Q_i$. If $U_i=\emptyset$ then $G_v$ has no e.d. - STOP. From now on, let $Q_i:=U_i$. If $U_i=\{d_i\}$ then $D_v:=D_v \cup \{d_i\}$. Delete all vertices $x \in N_3$ which have a neighbor and a non-neighbor in $N_4$. Contract $N_4$ to one vertex $z$ if $N_4 \neq \emptyset$.
\item[(3)] For all $|Q_i|=1$, add its vertex to $D_v$ and delete its neighbors from $N_2$. If there is an $i \in \{1,\ldots,\ell\}$ with $|Q_i|>1$, say $|Q_1|>1$, then check whether there is a vertex $d \in Q_1$ which has exactly the remaining $N_2$ vertices as its neighborhood in $N_2$ and sees $z$ for $N_4=\{z\}$.   
\item[(4)] Finally check whether $D_v$ is an e.d. of $G_v$ - if not then either $G$ is not $(P_6,S_{1,2,2})$-free or has no e.d. (containing $v$) of finite weight.
\end{enumerate}

\begin{theorem}\label{correcttime} 
Algorithm $(P_6,S_{1,2,2})$-Free-WED-$G_v$ is correct and runs in time $O(m)$.
\end{theorem}
 
\noindent 
{\bf Proof.} 
{\em Correctness.} The correctness follows from the previous claims and considerations.

\medskip
 
\noindent 
{\em Time bound.} The linear time bound is obvious.
\qed

\begin{corollary}\label{WEDP6S122fr} 
WED is solvable in time $O(\delta(G) m)$ for $(P_6,S_{1,2,2})$-free graphs.
\end{corollary}

\subsection{WED for $P_6$-free graphs of diameter 3}

In this subsection, we reduce the WED problem on $P_6$-free graphs in polynomial time to such graphs having diameter 3. Let $D$ be an e.d. of $G$. By Theorem~\ref{thm:prime}, we can assume that $G$ is prime. As before, we check for every vertex $v \in V$ if $v \in D$ leads to an e.d. of $G$. For this purpose, let $N_i$, $i \ge 1$, again be the distance levels of $v$. Recall that by (\ref{N5emptyN4stable}), $N_k = \emptyset$ for $k \ge 5$ and $N_4$ is an independent vertex set, and 
by (\ref{atmostoneDN3N4P6}), at most one vertex in $D_v \cap N_3$ has neighbors in $N_4$.

Recall that $A(x)$ denotes the anti-neighborhood of $x$. Thus, if $N_4 \neq \emptyset$ then check for every vertex $x \in N_3$ whether $\{v,x\} \cup (A(x) \cap N_4)$ is an e.d. in $G$; since $N_4$ is independent, vertices in $N_4$ not dominated by $x$ must be in $D_v$. This can be done in polynomial time for all $v$ with $N_4 \neq \emptyset$. 

Now we can assume that the diameter of $G$ is at most 3, i.e., for every $v \in V$, the distance level $N_4$ is empty. 

\begin{corollary}\label{WEDP6frdiam3} 
If WED is solvable in polynomial time for $P_6$-free graphs of diameter $3$ then WED is solvable in polynomial time for $P_6$-free graphs.
\end{corollary}

\section{WED for $(2P_3,S_{1,2,2})$-Free Graphs in Time $O(\delta(G) n^3)$}\label{2P3S122free}

In this section we improve the time bound $O(n^5)$ for WED \cite{BraMilNev2013} to $O(\delta(G) n^3)$ and simplify the proof in  \cite{BraMilNev2013}.  
Let $G=(V,E)$ be a connected $(2P_3,S_{1,2,2})$-free graph, let $v \in V$ and let $N_1,N_2,\dots$ be the distance levels of $v$. Since $G$ is $2P_3$-free, we have $N_k = \emptyset$ for $k \ge 6$. Let $R:=V \setminus (\{v\} \cup N_1 \cup N_2)$. Assume that $G$ admits an e.d.\ $D_v$ of finite weight with $v \in D_v$. 
Let $G_v:=G[N_2 \cup N_3 \cup N_4 \cup N_5]$, i.e. $G_v=G[N_2 \cup R]$; we can assume that $G_v$ is prime. Since $D_v$ is an e.d., $R \neq \emptyset$.  
Let $Q_1,\ldots,Q_{\ell}$, $\ell \ge 1$, denote the connected components of $G[R]$. Clearly, $D_v \cap Q_i \neq \emptyset$ for every $i$. 
Let $D_v \setminus \{v\}=\{d_1,\ldots,d_k\}$, and assume that $k \ge 2$ (otherwise, $G_v$ would have a universal vertex which is impossible for a prime graph). 
Since $G$ is $S_{1,2,2}$-free and $D_v$ is an e.d., we have: 
\begin{equation}\label{N2seemiss}
\mbox{ Every } x \in N_2 \mbox{ seeing a vertex } d_i \in D_v \mbox{ misses } N[d_j] \cap R, j \neq i.
\end{equation}

We claim: 
\begin{equation}\label{Ndicliques}
\mbox{ For every } i=1,\ldots,k, N[d_i] \cap R \mbox{ is a clique.}
\end{equation}
{\em Proof.}
Suppose that $N[d_1] \cap R$ is not a clique, i.e., there are neighbors $x,y \in R$ of $d_1$ with $xy \notin E$. Let $b \in N_2$ be a neighbor of $d_2$. By (\ref{N2seemiss}), $b$ misses $x$ and $y$ but now, $a,b,d_2,x,d_1,y$ induce $2P_3$, a contradiction.
\qed

Next we claim:
\begin{equation}\label{kge3N3disjunionofcliques}
\mbox{ If } k \ge 3 \mbox{ then } G[N_3] \mbox{ is the disjoint union of cliques.}
\end{equation}

\noindent
{\em Proof.}
Suppose that $k \ge 3$ and there is an edge $uw \in E$ for $u \in N(d_2) \cap R$ and $w \in N(d_3) \cap R$.
Let $x \in N_2$ with $xd_1 \in E$ and $a \in N_1$ with $ax \in E$.
Then by (\ref{N2seemiss}), $v,a,x,d_2,u,w$ induce $2P_3$, a contradiction.
\qed

\medskip

Thus, for $k \ge 3$, every $Q_i$ is a clique containing exactly one $D_v$ vertex: 
\begin{equation}\label{DQi}
|D_v \cap Q_i| = 1.
\end{equation}

If $Q_i$ is a single vertex $q_i$ then $q_i$ is forced and has to be added to $D_v$. From now on assume that for all $i$, $|Q_i| \ge 2$. By (\ref{N2seemiss}), we have:
\begin{equation}\label{N2sQimQj}
\mbox{If } z \in N_2 \mbox{ sees } Q_i \mbox{ then it misses all } Q_j, j \neq i.
\end{equation}

Let $S_i$ denote the set of vertices in $N_2$ distinguishing vertices in $Q_i$. Since $Q_i$ is not a module, $S_i \neq \emptyset$ for all $i \in \{1,\ldots,k\}$.   
Let $U_i$ denote the vertices in $Q_i$ which have a join to $S_i$; $U_i \neq \emptyset$ since $S_i$ vertices must have a $D_v$ neighbor in $Q_i$. 
We claim:
\begin{equation}\label{Uisize1}
\mbox{For all } i \in \{1,\ldots,k\}, |U_i|=1.
\end{equation}

\noindent
{\em Proof.}
Assume to the contrary that $|U_1|>1$. If $x \in S_1$ then by (\ref{N2sQimQj}), $xd_1 \in E$, i.e., $d_1 \in U_1$. Now, a vertex distinguishing $U_1$ would be in $S_1$ but vertices in $U_1$ have a join to $S_1$ and thus cannot be distinguished which is a contradiction to the assumption that $G_v$ is prime.      
\qed

\medskip

The other case when $k \le 2$, i.e., $|D_v \setminus \{v\}| \le 2$, can be easily done via the adjacency matrix of $G$: For any pair $x,y \in R$, $x \neq y$, with $xy \notin E$, check whether all other vertices in $G_v$ are adjacent to exactly one of them; this can be done in time $O(n^3)$. 

\medskip   

This leads to the following: 

\medskip

{\bf Algorithm $(2P_3,S_{1,2,2})$-Free-WED-$G_v$:} 

\medskip

\noindent
{\bf Given:} Connected graph $G=(V,E)$ and prime graph $G_v=G[N_2 \cup R]$ as constructed above with vertex weights $w(x)$; for all $x \in N_2$, $w(x)=\infty$. \\ 
{\bf Output:} An e.d. $D_v$ of $G_v$ of finite weight if $G_v$ has such an e.d., or the statement that $G$ is not $(2P_3,S_{1,2,2})$-free or $G_v$ does not have any e.d. of finite weight.

\begin{enumerate}
\item[(0)] Initially, $D_v:=\emptyset$.

\item[(1)] Determine $N_1,N_2$ and $R$. If $R=\emptyset$ then $G_v=G[N_2 \cup R]$ has no e.d. - STOP. Else determine the connected components $Q_1,\ldots,Q_{\ell}$ of $R$.  

\item[(2)] If $G[R]$ is not the disjoint union of cliques $Q_1,\ldots,Q_{\ell}$, $\ell \ge 3$, then check whether $G_v$ has a finite weight e.d. with two vertices, and determine an e.d. with minimum weight. If not, $G_v$ has no e.d. of finite weight - STOP.   

\item[(3)] (Now $G[R]$ is the disjoint union of cliques $Q_1,\ldots,Q_{\ell}$, $\ell \ge 3$) If $Q_i = \{d_i\}$ then $d_i$ is forced - $D_v:=D_v \cup \{d_i\}$. 
If $|Q_i| > 1$ then determine the set $S_i$ of vertices distinguishing $Q_i$, and determine the set $U_i$ of vertices in $Q_i$ having a join to $S_i$. 
If $U_i=\emptyset$ then $G_v$ has no e.d. - STOP. Otherwise, $U_i = \{d_i\}$ and $d_i$ is forced - $D_v:=D_v \cup \{d_i\}$.   

\item[(4)] Finally check whether $D_v$ is an e.d. of finite weight of $G_v$ - if not then either $G$ is not $(2P_3,S_{1,2,2})$-free or has no e.d. of finite weight.
\end{enumerate}

\begin{theorem}\label{2P3S122correctnesstime} 
Algorithm $(2P_3,S_{1,2,2})$-Free-WED-$G_v$ is correct and runs in time $O(n^3)$.
\end{theorem}
  
\noindent
{\bf Proof.} {\em Correctness.} The correctness follows from the previous claims and considerations.

\medskip

\noindent 
{\em Time bound.} The time bound is obvious since step (1) can be done in time $O(m)$, step (2) can be done in time $O(n^3)$, and steps (3) and (4) can be done in time $O(m)$.
\qed

\begin{corollary}\label{WED2P3S122fr} 
WED is solvable in time $O(\delta(G) n^3)$ for $(2P_3,S_{1,2,2})$-free graphs.
\end{corollary}

\noindent 
{\bf Acknowledgement.} 
The authors thank Martin Milani\v c for various helpful comments and the cooperation on the topic of Efficient Domination. 

\begin{footnotesize}

\end{footnotesize}


\begin{thebibliography}{99}

\bibitem{BalYu1989}
  E. Balas and C.S. Yu,
  On graphs with polynomially solvable maximum-weight clique problem,
  {\sl Networks} 19.2 (1989) 247-253.
%%%ok

%\bibitem{BanBarSla1988}
%    D.W. Bange, A.E. Barkauskas, P.J. Slater,
%    Efficient dominating sets in graphs,
%    in: R.D. Ringeisen and F.S. Roberts, eds., Applications of Discrete Math. (SIAM, Philadelphia, 1988) 189--199.

%\bibitem{BanBarHosSla1996}
%    D.W. Bange, A.E. Barkauskas, L.H. Host, P.J. Slater,
%    Generalized domination and efficient domination in graphs,
%    {\sl Discrete Math.} 159 (1996) 1-11.

%\bibitem{Biggs1973}
%   N.~Biggs,
%   Perfect codes in graphs,
%   {\sl J. of Combinatorial Theory (B)}, 15 (1973) 289-296.

%\bibitem{BraFicLeiMil2013}
%  A. Brandst{\"a}dt, P. Fi{\v c}ur, A. Leitert and Martin Milani{\v c},
%  Polynomial-time Algorithms for Weighted Efficient Domination Problems in AT-free Graphs and Dually Chordal Graphs,
%  Manuscript (2013)

%\bibitem{BraHunNev2010}
%    A. Brandst\"adt, C. Hundt, R. Nevries,
%    Efficient Edge Domination on Hole-Free graphs in Polynomial Time,
%    Conference Proceedings LATIN 2010, LNCS 6034, (2010) 650-661.

%\bibitem{BraLeiRau2012}
%    A. Brandst\"adt, A. Leitert, D. Rautenbach,
%    Efficient Dominating and Edge Dominating Sets for Graphs and Hypergraphs,
%    extended abstract in: Conference Proceedings of ISAAC 2012, LNCS 7676, 2012, 267-277.

\bibitem{BraMilNev2013}
    A. Brandst\"adt, M. Milani\v c, and R. Nevries, 
    New polynomial cases of the weighted efficient domination problem, 
    extended abstract in: Conference Proceedings of MFCS 2013, LNCS 8087, 195-206; full version: arXiv:1304.6255v1.

%\bibitem{BraMos2011}
%    A. Brandst\"adt, R. Mosca,
%    Dominating induced matchings for $P_7$-free graphs in linear time,
%    extended abstract in: T. Asano et al. (Eds.): Proceedings of ISAAC 2011, LNCS 7074, 100-109, 2011. Available online in {\sl Algorithmica} 2012.

%\bibitem{BreCorHabPau2008}
%    A. Bretscher, D.G. Corneil, M. Habib, Ch. Paul,
%    A Simple Linear Time LexBFS Cograph Recognition Algorithm,
%    {\sl SIAM J. Discrete Math.} 22(4) (2008) 1277-1296.

%\bibitem{CarKorLoz2011}
%     D.M.~Cardozo, N. Korpelainen, V.V.~Lozin,
%     On the complexity of the dominating induced matching problem in hereditary classes of graphs,
%     {\sl Discrete Applied Math.} 159 (2011) 521-531.

%\bibitem{Chang1997}
%M.-S. Chang, Weighted domination of cocomparability graphs, {\sl Discrete Applied Math.} 80 (1997)
%135-148.

%\bibitem{ChaLiu1993}
%    M.-S. Chang, Y.-C. Liu,
%    Polynomial algorithms for the weighted perfect domination problems on chordal graphs and split graphs,
%    {\sl Information Processing Letters} 48 (1993) 205-210.

%\bibitem{ChaLiu1994}
%M-S. Chang and Y.-C. Liu,
%Polynomial algorithms for weighted perfect domination problems on interval and circular-arc graphs,
%{\sl J. Inf. Sci. Eng. 11} (1994) 549-568.

%\bibitem{ChaPanCoo1995}
%    G.J. Chang, C. Pandu Rangan, S.R. Coorg,
%    Weighted independent perfect domination on co-comparability graphs,
%    {\sl Discrete Applied Math.} 63 (1995) 215-222.

%\bibitem{CorPerSte1985}
%    D.G. Corneil, Y. Perl, L.K. Stewart,
%    A linear recognition algorithm for cographs,
%    {\sl SIAM J. Computing} 14 (1985) 926-934.

%\bibitem{CouMakRot2000}
%   B. Courcelle, J.A. Makowsky and U. Rotics,
%   Linear time solvable optimization problems on graphs of bounded clique width,
%   {\sl Theory of Computing Systems} {\bf 33} (2000) 125-150.

%\bibitem{EscWan2014}
%  E.M.\ Eschen and X.\ Wang,
%  Algorithms for unipolar and generalized split graphs,
%  {\sl Discrete Applied Mathematics} 162 (2014) 195-201.

\bibitem{Farbe1989}
  M. Farber,
  On diameters and radii of bridged graphs,
  {\sl Discrete Mathematics}, 73.3 (1989) 249-260.
%%%ok

%\bibitem{FelHoo2000}
%   M.R. Fellows, M.N. Hoover,
%   Perfect Domination,
%   {\sl Australasian J. of Combinatorics} 3 (1991) 141-150.

%\bibitem{FoeHam1977}
%    S. F\"oldes, P.L. Hammer,
%    Split graphs,
%    {\sl Congressus Numerantium} 19 (1977) 311-315.

%\bibitem{GarJoh1979}
%    M.R. Garey, D.S. Johnson,
%    Computers and Intractability -- A Guide to the Theory of NP-completeness,
%    {\sl Freeman}, San Francisco, 1979.

%\bibitem{GriSlaSheHol1993}
%    D.L. Grinstead, P.L. Slater, N.A. Sherwani, N.D. Holmes,
%    Efficient edge domination problems in graphs,
%    {\sl Information Processing Letters} 48 (1993) 221-228.

%\bibitem{Karp1972}
%    R.M. Karp,
%    Reducibility among combinatorial problems,
%    In: {\em Complexity of Computer Computations}, Plenum Press, New York (1972) 85-103.

%\bibitem{Krato1991}
%J. Kratochv\'il, Perfect codes in general graphs, Rozpravy \v Ceskoslovensk\'e Akad. V\v ed \v Rada
%Mat. P\v r\'irod V\v d 7 (Akademia, Praha, 1991).

%\bibitem{Laroc1992}
% P. Laroche,
% Planar 1-in-3 satisfiability is NP-complete,
% ASMICS Workshop on Tilings, Deuxi\`eme Journ\'ees Polyominos et pavages,
% Ecole Normale Sup\'erieure de Lyon, 1992.

%\bibitem{Leite2012}
%    A. Leitert,
%    Das Dominating Induced Matching Problem f\"ur azyklische Hypergraphen,
%    Diploma Thesis, University of Rostock, Germany, 2012.

%\bibitem{LiaLuTan1997}
%   Y.D. Liang, C.L. Lu, C.Y. Tang,
%   Efficient domination on permutation graphs and trapezoid graphs,
%   in: {\sl Proceedings COCOON'97}, T. Jiang and D.T. Lee, eds., {\sl Lecture Notes in Computer Science} Vol. 1276
%   (1997) 232-241.

%\bibitem{Lin1998}
%   Y.-L. Lin,
%   Fast algorithms for independent domination and efficient domination in trapezoid graphs,
%   in: {\sl Proceedings ISAAC'98}, {\sl Lecture Notes in Computer Science} Vol. 1533
%   (1998) 267-275.

%\bibitem{LivSto1988}
%   M. Livingston, Q. Stout,
%   Distributing resources in hypercube computers,
%   in: {\sl Proceedings 3rd Conf. on Hypercube Concurrent Computers and Applications} (1988) 222-231.

%\bibitem{LozMos2009}
%    V.V. Lozin, R. Mosca,
%    Maximum independent sets in subclasses of $P_5$-free graphs,
%    {\sl Information Processing Letters} 109 (2009) 319-324.

%\bibitem{LozMos2012}
%    V.V. Lozin, R. Mosca,
%    Maximum regular induced subgraphs in $2P_3$-free graphs,
%    {\sl Theoretical Computer Science} 460 (2012) 26-33.

%\bibitem{LuTan1998}
%    C.L. Lu, C.Y. Tang, Solving the weighted efficient edge domination problem on bipartite permutation graphs,
%   {\sl Discrete Applied Math.} 87 (1998) 203-211.

%\bibitem{LuTan2002}
%    C.L. Lu, C.Y. Tang, Weighted efficient domination problem on some perfect graphs,
%   {\sl Discrete Applied Math.} 117 (2002) 163--182.

%\bibitem{LuKoTan2002}
%    C.L. Lu, M.-T. Ko, C.Y. Tang,
%    Perfect edge domination and efficient edge domination in graphs,
%{\sl Discrete Applied Math.} 119 (2002) 227-250.

\bibitem{McCSpi1999}
    R.M. McConnell, J.P. Spinrad,
    Modular decomposition and transitive orientation,
    {\sl Discrete Math.} 201 (1999) 189-241.
%%%ok

%\bibitem{Milan2012}
%    M. Milani\v c,
%    Hereditary Efficiently Dominatable Graphs,
%    available online in: {\sl Journal of Graph Theory} 2012.

%\bibitem{MooRob2001}
% C. Moore, J.M. Robson,
% Hard Tiling Problems with Simple Tiles,
% \emph{Discrete \& Computational Geometry} 26 (2001) 573--590.

%\bibitem{MulRot2008}
% W. Mulzer, G. Rote,
% Minimum-Weight Triangulation is NP-hard,
% \emph{J. ACM} 55 (2008), Article No. 11.

\bibitem{Nevri2014}
R. Nevries,
Efficient Domination and Polarity,
Ph.D.~Thesis, University of Rostock, 2014.
%%%ok

\bibitem{Prisn1995}
  E. Prisner,
  Graphs with few cliques, in: Graph Theory, Combinatorics, and Applications: Proceedings of 7th Quadrennial International Conference on the Theory and Applications of Graphs (Y. Alavi, A. Schwenk ed.) John Wiley and Sons, Inc. (1995)
  945-956, New York, Wiley, 1995.
%%%ok

%\bibitem{RanSch2010}
%  B. Randerath, I. Schiermeyer,
%  On maximum independent sets in $P_5$-free graphs,
%  \textsl{Discrete Applied Math.} 158 (2010), 1041-1044.

\bibitem{Spinr2003}
    J.P. Spinrad,
    Efficient Graph Representations,
    Fields Institute Monographs, American Math. Society, 2003.

\bibitem{TsuIdeAriShi1977}
  S. Tsukiyama, M. Ide, H. Ariyoshi and I. Shirakawa,
  A New Algorithm for Generating All the Maximal Independent Sets,
  {\sl SIAM J. Comput.} 6.3 (1977) 505-517.

%\bibitem{Yen1992}
%    C.-C. Yen,
%    Algorithmic aspects of perfect domination,
%    Ph.D. Thesis, Institute of Information Science, National Tsing Hua University, Taiwan 1992.

%\bibitem{YenLee1996}
%    C.-C. Yen, R.C.T. Lee,
%    The weighted perfect domination problem and its variants,
%{\sl Discrete Applied Math.} 66 (1996) 147-160.

\end{thebibliography}
\end{document}